\documentclass[a4paper,12pt]{article}

\usepackage[dvips]{graphicx}
\usepackage{amsmath}

\title{Nonlinear Switching Dynamics in a Nanomechanical Resonator}

\author{Quirin P. Unterreithmeier, Thomas Faust and J\"org P. Kotthaus}

\begin{document}

\maketitle

\begin{center} \normalsize{Fakult{\"a}t f{\"u}r Physik and Center
for NanoScience (CeNS), Ludwig-Maximilians-Universit{\"a}t,\\
Geschwister-Scholl-Platz 1, D-80539 M{\"u}nchen, Germany}
\end{center}

\begin{abstract}
{\bf
The oscillatory response of nonlinear systems exhibits characteristic phenomena such as multistability\,\cite{basins}, discontinuous jumps\,\cite{parallel_readout,relativistic_electron} and hysteresis\,\cite{relativistic_electron}. These can be utilized in applications leading, e.g., to precise frequency measurement\,\cite{noise_enabled_prec}, mixing\,\cite{mixing}, memory elements\,\cite{parametron, memory_el}, reduced noise characteristics in an oscillator\,\cite{osc_noise_evasion} or signal amplification\,\cite{sr_signal_amp,sr_coh_signal_amp,josephson_bifurcation_amp,noise_squeezing}. Approaching the quantum regime\cite{evanes_coupl}, concepts have been proposed that enable low backaction measurement techniques\,\cite{josephson_bifurcation_amp} or facilitate the visualisation of quantum mechanical effects\,\cite{quant_trans}. Here we study the dynamic response of nanoelectromechanical resonators in the nonlinear regime aiming at a more detailed understanding and an exploitation for switching applications. Whereas most previous investigations concentrated on dynamic phenomena arising at the onset of bistability\,\cite{basins,noise_enabled_prec,sr_signal_amp}, we present experiments that yield insight into the non-adiabatic evolution of the system while subjected to strong driving pulses and the subsequent relaxation. Modeling the behaviour quantitatively with a Duffing oscillator, we can control switching between its two stable states at high speeds, exceeding recently demonstrated results by $10^5$\,\cite{parametron,memory_el}.
}
\end{abstract}

Nano-Electro-Mechanical Systems (NEMS) have been established as excellent devices to explore nonlinear dynamical behaviour, as they exhibit high mechanical quality (Q) factors\,\cite{million_Q, CNT_res}, fast response times\,\cite{GHz_resonator}, fairly low drift\,\cite{basins} and can be easily excited into the nonlinear regime\,\cite{basins}. Yet most systematic studies of the dynamics of such systems concentrated on the regime near the onset of bistability\,\cite{basins,noise_enabled_prec}. Complementary, we explore the response of a driven nonlinear nanomechanical resonator to excitation by intense and short radio frequency (RF) pulses that drive the resonator away from the stationary points. The observed response is found to be in excellent agreement with simulations based on a Duffing oscillator\,\cite{nonlin_osc}. Therefore we can make use of pulsed excitations to controllably switch the resonator between the stable points on a time scale that is no longer limited by the relaxation time of the system\,\cite{parametron,memory_el}.

The resonator consists of a doubly-clamped silicon nitride string of dimensions $35\,\mu\rm{m} \cdot 250\,\rm{nm} \cdot 100\,\rm{nm}$ (length, width, height, respectively) under high tensile stress, leading to high mechanical Q-factors\,\cite{universal_trans}. In vacuum and at room temperature we electrically excite the resonator at RF frequencies employing dielectric gradient forces provided by suitably located and biased electrodes\,\cite{universal_trans,polarization_force}. Illuminating the resonator with a light emitting diode, we detect the resonant motion by a small on-chip Schottky diode fabricated close to the resonator and serving as a photodetector for the oscillating component of the optical near-field as will be discussed in detail elsewhere\,\cite{on-chip_det}. The nonlinear resonator is continuously actuated by the RF output of a network analyzer as depicted in Fig.\,1a. Applying sufficiently strong excitation amplitudes, the mechanical response around resonance tends to bend towards higher frequencies as depicted in Fig.\,1b, corresponding to string-hardening. This behaviour can be quantitatively modeled by solving the so-called Duffing equation\,\cite{nonlin_osc}, an extension of the simple harmonic oscillator by a nonlinear term of 3rd order.
\begin{equation}
  \label{Duff_eq}
  \ddot{x}(t) + \frac{2 \pi f_0}{Q} \dot{x}(t) + (2 \pi f_0)^2 x(t) + \alpha_3 x(t)^3 = k \cos(2 \pi(f_0 + \sigma)t)
\end{equation}
Here, $x(t)$ designates resonator displacement, $f_0 = 8\,{\rm MHz},\,Q=1.2\cdot 10^5$ its resonance frequency and quality factor; $\alpha_3$ is the cubic correction to the linear restoring force. The excitation amplitude is $k$ and its frequency-detuning from the mechanical resonance is $\sigma = f-f_0$. We apply a perturbation calculation using the ansatz $x(t) = a(t) \cos(2\pi(f_0 + \sigma) t + \gamma(t))$, with time-dependent displacement amplitude $a(t)$ and phase $\gamma(t)$. This leads to the two coupled eqs.\,\cite{nonlin_osc}:
\begin{align}
  \label{time_evol}
  \dot{a}(t) &= - \frac{2 \pi f_0 a(t)}{2Q}+\frac{k \sin(\gamma(t))}{4 \pi f_0}\nonumber \\
  \dot{\gamma}(t) &= 2 \pi \sigma - \frac{3\alpha_3 a(t)^2}{16 \pi f_0}+\frac{k \cos(\gamma(t))}{4\pi f_0 a(t)}
\end{align}
By setting $\dot{a}(t)=0, \dot{\gamma}(t)=0$, one arrives at the quasi-static solution $a=a(f)$. This curve can be excellently fitted to the measured data (see Fig.\,1b), thereby obtaining the numerical values for $f_0$, $Q$ and $\alpha_3$. In the following, all displacements $a(t)$ are given in units normalized to the critical displacement $a_c$. This critical displacement marks the onset of bistability, at which the first and second derivative with respect to $f$ vanish ($\partial a/\partial f = 0, \partial^2 a/(\partial f)^2 =0$). Here, the critical displacement corresponds to a half-peak-to-peak displacement of approximately 6\,nm.

Figure\,1c shows the calculated displacement response of the resonator when actuated with an excitation amplitude that is ten times larger than the critical actuation leading to the critical displacement $a_c$. In the following, we always continuously excite our system $\sigma = 1\,{\rm kHz}$ above resonance, well in the bistable regime. The two stable oscillatory amplitudes are marked as blue dots in Fig.\,1c.

To gain insight into the dynamical behaviour of our system, we measure the relaxation towards one of these stable points of the continuously driven string after additional pulsed excitation. The pulsed excitation is provided by the output of a frequency generator that is phase-locked to the network analyzer and operates at the same frequency $f$ as the continuous drive. The phase of the frequency generator's signal can be adjusted to any phase $\varphi$ with respect to the continuous drive as sketched in Fig.\,1a. To avoid confusion, we always specify the two phases with their respective symbol $\gamma$ or $\varphi$ in the ongoing text. An RF switch serves to define RF pulses of adjustable duration.

Any non-stationary resonator state (defined by its displacement amplitude $a$ and phase $\gamma$ referred to the continuous drive; or equivalently by its in ($a \cos(\gamma)$) and out-of-phase ($a \sin(\gamma)$) amplitude component) will converge towards either of the two stable states. This convergence divides the resonator's phase space into two basins of attraction\,\cite{basins,nonlin_osc}, as depicted in Fig.\,2a as black and white regions, obtained by numerically integrating eqs.\,\ref{time_evol} (see Methods). To test this simulated behaviour experimentally, we apply the described short and intense RF pulse that excites the oscillator away from the stable state. Immediately after switching off this pulse, we start recording the resonator state with a sampling rate of 100\,kHz. Depending on amplitude, duration and phase $\varphi$ of the pulsed excitation, the resonator's dynamic state starts in either the white or black region of phase space directly after excitation and relaxes in a spiraling motion towards either of the stable states staying within the respective region of phase space as theoretically predicted. In Fig.\,2a two measured traces differing in the phase $\varphi$ of the previously applied RF pulse are plotted in the phase space and show excellent agreement with theory. Figures\,2b,c display the evolution with time, showing fast dynamics for high amplitudes. Eventually, the state oscillates around either of the stable points with constant frequency.

We intend to utilize the pulses to controllably switch between the stable points, therefore we apply an indirect measurement scheme to explore the non-adiabatic time evolution during strong pulse excitation. Such an indirect scheme is needed because electric crosstalk prevents a direct measurement of the resonator's state during the strong RF pulses. Therefore, by varying the length of the excitation pulses at a given actuation amplitude, we identify the consecutive crossings through the two basins of attraction as indicated in Fig\,3a. The corresponding trace is obtained by time-integrating eqs.\,\ref{time_evol}, taking the additional excitation into account. In the experiment, the oscillator is prepared in its lower stable state by subsequently switching off and on the continuous actuation. We apply a short RF pulse with an excitation amplitude that is always 18 times larger than the continuous actuation. After waiting several relaxation times given by $Q/(2 \pi f_0)$, the attained stable state is recorded in displacement amplitude $a$ and phase $\gamma$. By systematically varying the length of the RF pulse and its phase $\varphi$, we implicitly map the end-point of the trace as shown in Fig.\,3a and thereby obtain the spirals in Fig.\,3b. This measured result is in excellent agreement with the calculation shown in Fig.\,3c, employing no fit parameters. The range of achieved displacement amplitudes extends those of previous measurements~\cite{basins} to values of ten times the critical amplitude $a_c$. We can deduce that the perturbation solution describing the time evolution eqs.\,\ref{time_evol} remains accurate at least up to displacements that correspond to ten times the critical amplitude $a_c$. This may prove useful whenever dynamics of nonlinear systems are seen in experiments\,\cite{laser_ramp,bec_dyn}.

Our quantitative understanding of the experiment enables us to predict the parameters needed in order to switch directly between the two stable states. We thus extend previous concepts\,\cite{parametron,memory_el} of switching limited by the relaxation time scale $Q/(2\pi f_0)$ to active switching via RF pulses suitably chosen in amplitude, phase and length. Figure\,4a shows two consecutive switching events; for the duration of the RF pulses of about $80\,\mu\rm s$ electric crosstalk produces overshoots partially exceeding the displayed range of displacement amplitudes. The nearly constant amplitude values highlighted by gray areas reflect the respective stable state of the bistable system. Note that the approach towards these constant amplitudes occurs on a time scale of less than 1\,ms and only reflects the limited dynamics of the electronic measurement setup in contrast to the mechanical relaxation behaviour studied in Fig.\,2b and c.

Since we can pulse towards either of the targeted stable states with high precision in phase space, we can switch between the stable states with a high repetition rate. Any systematic deviation would add up, eventually preventing controllable switching. Figure\,4b shows ten consecutive switching events within ten milliseconds each going back and forth between the two stable states. This corresponds to a demonstrated operating speed of 2\,kHz. The applied pulse duration of $80\,\mu$s of a single pulse allows operation speeds of approximately 11\,kHz. In Fig.\,4c, we plot the same switching sequence in phase space. The image shows some systematic deviation of the measured traces with respect to the predicted stable states occurring immediately after the application of an RF pulse. This deviation is a result of the electric crosstalk and the finite bandwidth of the measurement setup. The experimentally chosen pulse durations deviate by less than 4\% from the ones that were predicted theoretically.

The operating speed is given by the resonance frequency divided by the number of oscillations required for one switching operation. At present, nanomechanical resonators reaching GHz resonance frequencies have been demonstrated\,\cite{GHz_resonator}. Increasing the applied RF pulse amplitude by two orders of magnitude, this would result in a memory element operating at 100\,MHz, approaching the speed of DRAMs. Given recent advances on the quality factor of high frequency resonators\,\cite{CNT_res}, the power consumption of such elements could be principally lower than CMOS-based memories\,\cite{parametron}.

In conclusion, we quantitatively study the dynamical oscillatory response of a nonlinear nanomechanical resonator in bistable configuration. The application of short RF pulses allows us to modify the resonator state at will. We utilize these pulses to highly excite the resonator. The measured results can be excellently modeled using a combination of perturbation calculation and numerical integration. We thereby directly confirm the accuracy of this model calculation to describe nonlinear dynamics\,\cite{quant_trans,bec_dyn}. Our quantitative understanding allows us to predict and generate RF pulse parameters that directly switch between the two stable states repeatedly. The duration of these switching pulses corresponds to approximately 1000 cycles of oscillation, significantly less than the number of oscillation required for the relaxation from an excited to a stable state given by the quality factor $Q = 1.2 \cdot 10^5$. The switching speed performance represents a five orders of magnitude improvement when compared to previous results\,\cite{parametron,memory_el}.

\subsubsection*{Methods}

The two simulated basins of attraction shown in Figs.\,2a,\,3a and 4c are obtained using the fact that any non-stationary oscillatory state of the nonlinear system theoretically converges towards either of the three stationary points, shown as the intersections of the vertical line and the calculated response in Fig.\,1c. The two intersections marked by blue dots correspond to the stable states. The state given by the third intersection is unstable; its basin of attraction is degenerate, forming a line in phase space (rather than an area). This line marks the boundary between the two basins of attraction and can be calculated by time-integrating eqs.\,\ref{time_evol} going backwards in time and starting in close proximity of the unstable stationary point\,\cite{basins}.

To obtain the time evolution during the application of an RF pulse, we integrate eqs.\,\ref{time_evol}, yielding a trace as a function of time as shown in Fig.\,3a. We then numerically determine the intersections of this trace with the boundary line between the basins of attraction, calculated as described above. Thereby we obtain a set of time values corresponding to pulse lengths that would lead exactly onto that boundary. This calculation is systematically repeated for different phases $\varphi$ of the applied pulse. We then plot these intersections as a function of pulse length and phase $\varphi$ in polar coordinates. It is straightforward to suitably join these points to form the "spiral" shown in Fig.\,3c. The turn of the rotational direction results from the fact that the absolute amplitude of all time traces such as the one displayed in Fig.\,3a reach their maximum at about $300\,\mu\rm s$ pulse duration and decrease afterwards.


\subsubsection*{Acknowledgements}
Financial support by the Deutsche Forschungsgemeinschaft via
project Ko 416/18 as well as the German Excellence Initiative
via the Nanosystems Initiative Munich (NIM) and LMUexcellent is gratefully
acknowledged. We thank E.~M. Weig for critically reading the manuscript.

\subsubsection*{Author Contributions}
Q.~P.~U. initiated the concept under the guidance of J.~P.~K. and performed the simulations; the fabrication and experiments have been carried out by T.~F. and Q.~P.~U. The results have been discussed and the manuscript has been written by all authors.

\subsubsection*{Author Information}
Reprints and permissions information is available at
npg.nature.com/reprintsandpermissions. The authors declare no
competing financial interests. Correspondence and requests for
materials should be addressed to quirin.unterreithmeier@physik.uni-muenchen.de

\newpage
\noindent {\bf Figure Captions}

\begin{figure}[h]
    \begin{center}
    \includegraphics{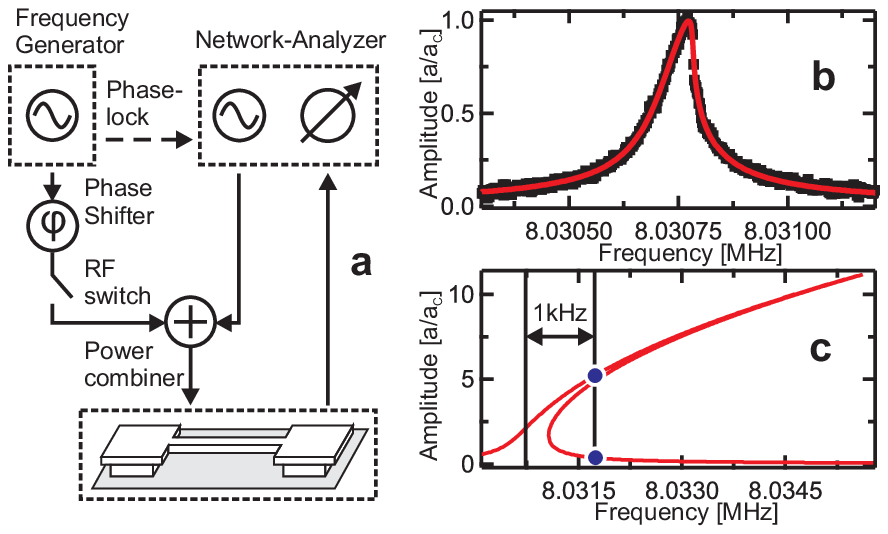}\\
    \end{center}
	\caption{}
\end{figure}

\noindent {\bf Fig.\,1. Setup and Quasistatic Response}  {\bf a} Schematic setup, a nanomechanical resonator is continuously actuated in the nonlinear regime using the RF output of a network analyzer; additional RF pulses of the same frequency are provided by a frequency generator that is phase-locked to the network analyzer; the output of the frequency generator can be adjusted to any phase $\varphi$ with respect to the continuous actuation; an RF switch defines short RF pulses. The resonator's oscillatory state, given by its displacement amplitude $a$ and phase $\gamma$, is recorded by a near-by photodiode and the network analyzer. {\bf b} Quasistatic response to continuous actuation near the onset of bistability, measurement (black) and fit (red) using a solution of the Duffing equation; the displacement amplitude $a$ is given in units of the critical displacement $a_c$, marking the onset of bistability. {\bf c} Calculated response when actuating ten times the critical driving amplitude. At an actuation frequency $f$, 1\,kHz above resonance $f_0$ two stable oscillation amplitudes exist, marked with blue dots; this actuation is used for all following measurements.

\newpage

\begin{figure}[h]
    \begin{center}
    \includegraphics{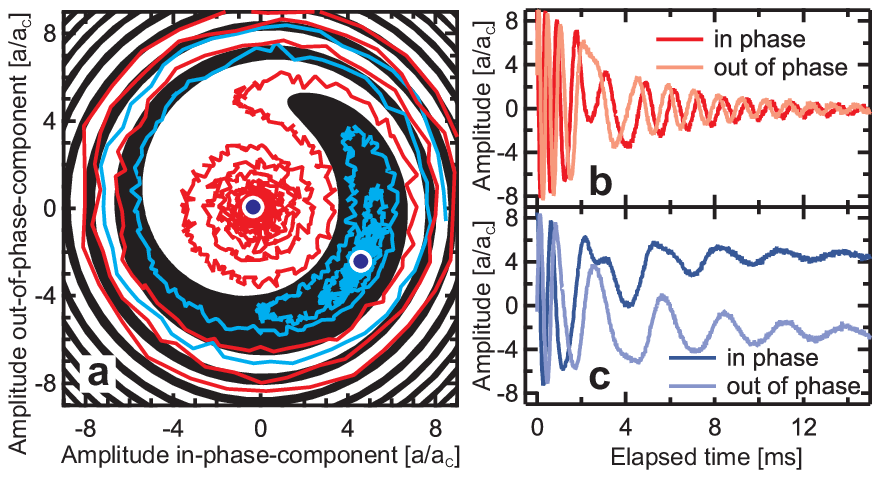}\\
    \end{center}
\end{figure}

\noindent {\bf Fig.\,2. Relaxation towards the stable points} {\bf a} The resonator's phase space is shown parametrised by the in ($a \cos(\gamma)$)and out-of-phase ($a \sin(\gamma)$) components of the oscillatory displacement. Because of the nonlinear actuation, two stable points exist (blue dots), each having its (calculated) basin of attraction (black/white: high/low displacement amplitude). The displayed traces show the measured relaxation of an oscillatory state after being excited to an amplitude $a \approx 8 a_c$ for two different excitation phase $\varphi$ settings. {\bf b} and {\bf c} display the same relaxation process versus time, the trace colour corresponds to {\bf a}.

\newpage

\begin{figure}[h]
    \begin{center}
    \includegraphics{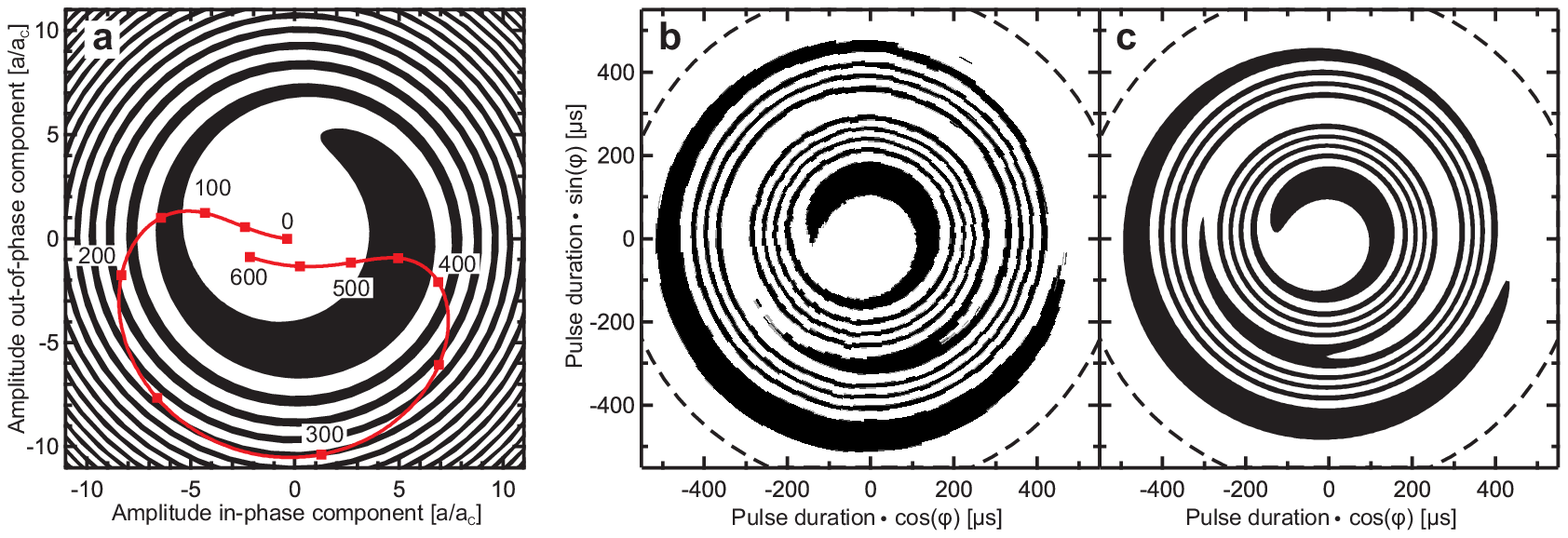}\\
    \end{center}
\end{figure}

\noindent {\bf Fig.\,3. Time evolution during the application of an RF pulse} {\bf a} Calculated trace (red) showing the time evolution in phase space subject to an RF pulse of phase $\varphi=-98^\circ$ and an amplitude 18-fold larger than the continuous actuation. The time trace starts from the lower stable state and crosses in alternating fashion the two basins of attraction which are the same as in Fig.\,2a; the displayed time values are given in $\mu$s. {\bf b} Measured final state (black/white: high/low displacement amplitude) after the application of such RF pulses systematically varied in duration and phase $\varphi$ and plotted in polar coordinates. {\bf c} Simulation of the measurement employing no fit parameters.

\newpage

\begin{figure}[h]
    \begin{center}
    \includegraphics{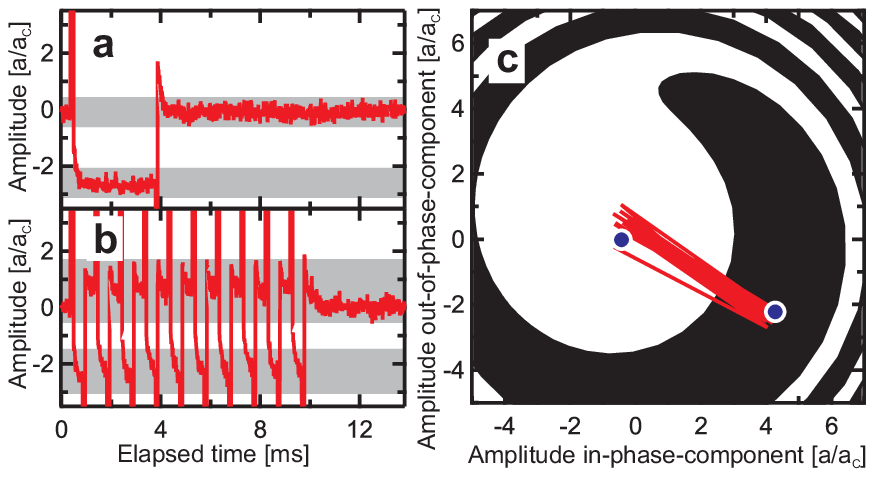}\\
    \end{center}
\end{figure}

\noindent {\bf Fig.\,4. Switching between the stable points} {\bf a} Out-of-phase component of the measured resonator displacement $a \sin(\gamma)$; the part of nearly constant amplitude (highlighted by the gray background) corresponds to the stable points; the spikes are a result of electric crosstalk when applying short RF pulses suitably chosen to directly switch between these states and do not correspond to displacement amplitudes. {\bf b} Consecutive switching; ten pairs of switching events are shown; the duration of one pulse is approximately $80\mu$s, the repetition rate of the pairs is 1\,kHz. Because of the finite measurement bandwidth and electric crosstalk there is a systematic deviation compared to {\bf a}. {\bf c} The same measurement displayed in phase space.

\newpage

\newpage

\end{document}